\documentclass{article}

\usepackage{arxiv}

\usepackage{amsmath}
\usepackage{comment}
\usepackage{amssymb}
\usepackage[utf8]{inputenc} 
\usepackage{graphicx}
\usepackage{subcaption}
\usepackage[T1]{fontenc}    
\usepackage{url}            
\usepackage{booktabs}       
\DeclareMathOperator*{\argmax}{argmax}
\DeclareMathOperator*{\argmin}{argmin}
\usepackage{algorithm}
\usepackage{algorithmic}
\usepackage{amsfonts}       
\usepackage{nicefrac}       
\usepackage{microtype}      
\usepackage{lipsum}
\usepackage{hyperref}
\newtheorem{theorem}{Theorem}[section]

\hypersetup{
    colorlinks=true,
    linkcolor=blue,
    filecolor=magenta,      
    urlcolor=cyan,
}

\title{Recursive Estimation for dynamical systems with measurement bias, outliers and constraints}

\author{
  Krishan Mohan Nagpal  
  \thanks{Krishan Nagpal is a Managing Director in Corporate Risk in Wells Fargo \& Co. The opinions expressed here are those of the author and do not represent those of his employer Wells Fargo \& Co.}\\
  Wells Fargo \& Co. \\
 \texttt{krishan$\_$nagpal@yahoo.com} \\
}

\begin{document}
\maketitle

\begin{abstract}
This paper describes recursive algorithms for state estimation of linear dynamical systems when measurements are noisy with unknown bias and/or outliers. For situations with noisy and biased measurements, algorithms are proposed that minimize $\epsilon$ insensitive loss function. In this approach which is often used in Support Vector Machines, small errors are ignored making the algorithm less sensitive to measurement bias. Apart from $\epsilon$ insensitive quadratic loss function, estimation algorithms are also presented for $\epsilon$ insensitive Huber M loss function which provides good performance in presence of both small noises as well as outliers. The advantage of Huber cost function based estimator in presence of outliers is due to the fact the error penalty function switches from quadratic to linear for errors beyond a certain threshold. For both objective functions, estimation algorithms are extended to cases when there are additional constraints on states and exogenous signals such as known range of some states or exogenous signals or measurement noises. Interestingly, the filtering algorithms are recursive and structurally similar to Kalman filter with the main difference being that the updates based on the new measurement ("innovation term") are based on solution of a quadratic optimization problem with linear constraints.
\end{abstract}

.

\section{Introduction}

The goal of this paper is to propose recursive Filtering algorithms for linear dynamical systems where measurements can be quite noisy with unknown bias and/or outliers. Kalman Bucy estimation framework, also sometimes described as $\mathcal{H}_2$ framework, provides optimal way to estimate states of the system when the power spectrum density of the noises and disturbances is known (see for example \cite{anderson} and \cite{meditch}). Kalman-Bucy approach also has certain optimal worst case properties. For example Krener (\cite{krener}) showed that Kalman-Bucy filter is a minimax filter with quadratic norm of exogenous signals.  

In some applications such as in finance, the data can be very noisy and unpredictable with unstable statistical properties of the exogenous signals/noises. For such situations when there is less information about noises, worst case approaches such as $\mathcal{H}_\infty$ have been proposed (see for example \cite{grimble}, \cite{nag1} and \cite{shaked}). 

 The $\epsilon$-insensitive loss function, one of the two loss functions considered in this paper, was introduced by Vapnik and coworkers (see for example \cite{vapnik1} and \cite{vapnik2}) for Support Vector Machine (SVM) algorithms in machine learning and regression. In $\epsilon$-insensitive SVM learning algorithms the cost function may be linear or quadratic and involve problem specific kernels but the common theme is that one ignores small errors which has been shown to provide more robustness and better generalizability of the algorithms. This framework also provides Bayesian estimate under some assumption with Gaussian noises with unknown nonzero means (\cite{pontil}). 

In presence of outliers, traditional estimation algorithms based on quadratic cost function can be overly influenced by outliers resulting in overall poor performance. In such cases estimation algorithms with Huber type cost functions (\cite{Huber}, \cite{durovic} and \cite{chan})  may be more appropriate as the cost function switches from quadratic to linear when errors become sufficiently large as is the case with outliers. The second cost function considered in this paper is a hybrid of Huber loss function and $\epsilon$-insensitive loss function as the one considered in \cite{rojo} for ARMA system identification. The proposed cost function a) ignores small measurement errors (is $\epsilon$-insensitive), and b) has linear instead of quadratic penalty for large errors making the estimates less sensitive to small noises as well as outliers. For linear dynamical systems, various other algorithms for handling outliers have also been proposed in \cite{aless}, \cite{andri} and \cite{garulli}. At a general level, Aravkin et al. \cite{aravkin} have considered estimation problems with such convex non-quadratic cost functions and drawn links to machine learning. 

There are instances when additional information is available about the system beyond the description of the the dynamical model. Examples of such additional information are maximum or minimum value of certain states (for example price of an asset can never be negative or physical constraints that limit movement of an object) or knowledge about the magnitude of disturbances and measurement noises such as an upper bound on measurement noise. Incorporating such additional information can be helpful not only in identification of outliers but can also lead to improved estimates as such information puts constraints on exogenous signals and possible trajectories of states. One approach for estimation under such constraints is to obtain sets constraining  possible state values (see for example \cite{milanese} and \cite{garulli}). At a more general overarching level, convex constraints with convex cost functions for smoothing problem have been considered by Aravkin et al. \cite{aravkin}. In this paper we assume the additional information about the system can be described in terms of inequality constraints that are linear with respect to states and exogenous signals and develop estimation algorithms that satisfy the constraints as well as minimize the two cost functions considered in this paper ($\epsilon$ insensitive quadratic as well as Huber).

In this paper we extend results from Nagpal \cite{nagpalarxiv} where optimal smoothing algorithms were developed for $\epsilon$-insensitive $\epsilon$ insensitive Huber M loss function. The algorithms in this paper apply those results for one step horizon to develop recursive algorithms that are easily implemented and where complexity does not grow with the number of observations. Remarkably, the algorithms bear strong structural resemblance to Kalman Filter with the primary difference being that the update based on the new information ("innovation term") is based on solution of a quadratic optimization problem with linear constraints. 

This paper is organized as follows. In the next section we describe the the two objective functions and the background results from Nagpal \cite{nagpalarxiv} which form the basis of the results presented here. Main results are described in Section 3 and the last section concludes with a summary of the results. 

\section{Problem Formulation and Background Results}

Throughout the paper, $N$ represents a positive integer which will be used to describe the number of measurements available for estimation. For vectors $v, w \in \mathbb{R}^n$, $v \geq w$ implies that all the components of the vector $v-w$ are non-negative. In particular for a real valued vector, $v \geq 0$ would imply that all elements of the vector $v$ are non-negative. For a matrix $C \in \mathbb{R}^{m \times n}$, $C'$ will indicate its transpose. For a vector $x_k \in \mathbb{R}^n$, $x_{k_j}$ denotes the $j'th$ element of $x_k$. 

For all the estimation problems we will assume that the underlying system is known and finite dimensional linear system of the following form: 

\begin{align}
\label{system}
    x_{k+1} & =  Ax_k+Bw_k \text{, initial condition $x_0$ is not known with $\bar{x}_0$ its best estimate} \nonumber \\
    y_k & =  Cx_k+v_k
\end{align}

where $x_k \in \mathbb{R}^n$ is the state, $y_k \in \mathbb{R}^m$ are the noisy measurements, $w_k$ and $v_k$ are unknown exogenous signals and measurement noises respectively. Given measurements $\{y_1, \hdots , y_N\}$, the filtering problem involves estimating $x_N$. The proposed algorithms described in this paper are applicable for linear time varying systems as well (when $A,B$ and $C$ depend on time index $k$ in equation (\ref{system})) but for ease of transparency, we will assume the system parameters $A \in \mathbb{R}^{n \times n}$, $B \in \mathbb{R}^{n \times l}$ and $C \in \mathbb{R}^{m \times n}$ are known constant matrices. 

For any $k \geq 0$, $\hat{x}_k$ will denote the estimate of $x_k$ based on the given measurements $\{y_1, \hdots , y_k\}$. $I_m$ will denote identity matrix of dimension $m$. Given a sequence of vectors $x_k$ and matrices $R_k$, we will use the following vector and diagonal matrix notation
\[
\text{vec}(\{x_k\}):=\begin{bmatrix} x_1 \\ \vdots \\ x_N \end{bmatrix} \; , \; 
\text{diag}(\{R_k\}):=\begin{bmatrix} R_1 & 0 & \hdots & 0 \\ 0 & R_2 & 0 & \hdots \\ \vdots & \vdots & \vdots & \vdots \\ 0 & \hdots & 0 & R_N \end{bmatrix} \;
\]

The two cost functions that are optimized in the estimation algorithms are $\epsilon$- insensitive quadratic loss function and $\epsilon$- insensitive Huber loss function and are illustrated in Figure 1 and described below. The loss functions are parameterized by user specified positive parameters $\epsilon$, $\kappa$ and $r$:
\begin{equation}
\label{epsiloncost}
    \epsilon \text{ insensitive quadratic loss function} \; \; \; \;  f_\epsilon(z;r,\epsilon)=
    \begin{cases}
    0 \; \text{  if  } \; |z| < \epsilon \\
   \frac{1}{2} r(|z|-\epsilon)^2 \; \text{  if  } \;  |z| \geq \epsilon 
    \end{cases}
\end{equation}

\begin{equation}
\label{hubercost}
    \epsilon \text{ insensitive Huber loss function} \; \; \; \; f_{\epsilon H}(z;r,\epsilon, \kappa)=
    \begin{cases}
    0 \; \text{  if  } \; |z| < \epsilon \\
   \frac{1}{2} r(|z|-\epsilon)^2 \; \text{  if  } \; \epsilon \leq |z| < \epsilon + \frac{\kappa}{r} \\
    \kappa(|z|-\epsilon-\frac{\kappa}{r})+\frac{\kappa^2}{2r}  \; \text{  if  } \;  |z| \geq \epsilon + \frac{\kappa}{r}
    \end{cases}
\end{equation}

In interpreting the cost functions, $z$ should be considered the prediction error  $(y_k-C\hat{x}_k)$. In both cases the cost function is zero for sufficiently small z (for $|z| \leq \epsilon$). The difference between the above two cost functions is for larger values of $z$ (when $|z| \geq \epsilon + \frac{\kappa}{r}$) - in this case the Huber cost function is linear rather than quadratic in $z$. This feature of linear as opposed to quadratic cost function for large errors, makes the algorithm based on Huber cost function less sensitive to outliers. 

The parameters and thresholds in the above cost functions are chosen so that the cost function is continuous at the switching points. Note that the above $\epsilon$- insensitive Huber loss function is not completely general as choices of quadratic and linear weights $r$ and $\kappa$ fix the level $\epsilon + \frac{\kappa}{r}$ where the function switches from quadratic to linear. The advantage of this functional form is that it makes the optimization problems more tractable while achieving the desired objective of linear penalty for large errors. If $\kappa$ is very large, the threshold level for switching from quadratic to linear ($\epsilon + \frac{\kappa}{r}$) would also be large and thus one would expect minimizing the above Huber cost function would lead to the same results as minimizing the $\epsilon$-insensitive quadratic loss function. 

\begin{figure}
    \begin{subfigure}{0.48\textwidth}
        \includegraphics[width=\textwidth]{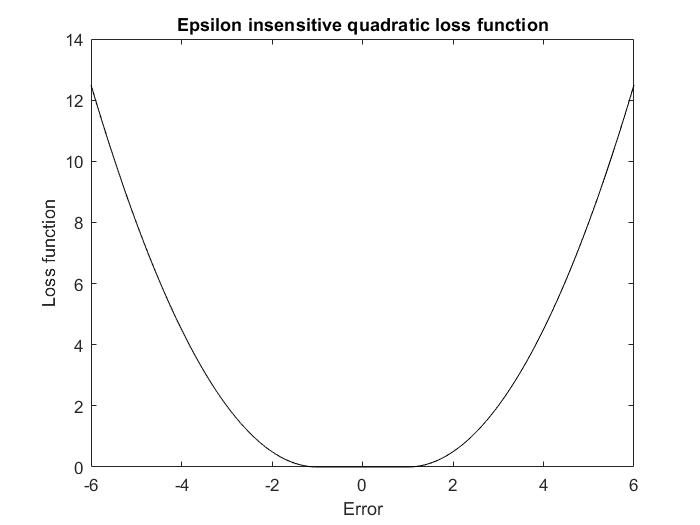}
        \caption{Quadratic $\epsilon$ insensitive loss function with $\epsilon=1$ and $r=1$}
    \end{subfigure} \hfill
    \begin{subfigure}{0.48\textwidth}
        \includegraphics[width=\textwidth]{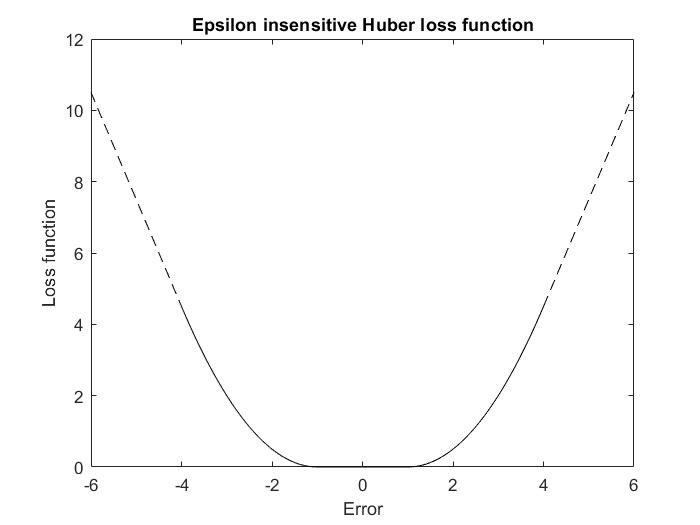}
        \caption{$\epsilon$ insensitive Huber loss function with $\epsilon=1$, $r=1$ and $\kappa=3$. The dashed line represents the region where the cost function is linear}
    \end{subfigure}
    \caption{Comparison of the two loss functions considered in this paper with $\epsilon=1$. Both loss functions are the same for $-4 \leq$Error$\leq4$ and both are zero for $-1 \leq$Error$\leq1$. For larger errors (when |Error|$>4$), Huber loss function (right plot) is linear and has smaller value than the quadratic loss function (left plot)}
 \end{figure}

We now describe the estimation problems and results described in Nagpal \cite{nagpalarxiv} that minimize the two cost functions described above. 

\textbf{Problem 1 -} \textit{Estimation with $\epsilon$- insensitive quadratic loss function :}
Let P, Q and R be given positive definite weighting matrices for initial conditions, disturbances and measurement noise. Given measurements $\{y_1, \hdots , y_N\}$ of the system (\ref{system}), consider the problem of estimating $\hat{x}_k$ from the following optimization problem :
\begin{equation}
    \label{problem1}
    \argmin_{\hat{x}_0,\hat{x}_k, \hat{w}_k, \eta_k}  \left[ (\hat{x}_0-\bar{x}_0)'P(\hat{x}_0-\bar{x}_0) + \sum_{k=0}^{N-1} \hat{w}_k'Q \hat{w}_k + \sum_{k=1}^{N} (y_k-C\hat{x}_k-\eta_k)'R(y_k-C\hat{x}_k-\eta_k)
    \right]
\end{equation}
subject to constraint that the estimated states match the given observations based on (\ref{system}) and the slack variable $\eta_k$ is less than $\epsilon$, i.e.
\begin{subequations}
\begin{align}
\label{constrdynamics}
    & \hat{x}_{k+1}  =  A\hat{x}_k+B\hat{w}_k \; , \; \forall \; k \text{ in } \{0,\cdots,N\} \\
    \label{constr2b}
   & | \eta_k |  \leq \epsilon \text{ where } \epsilon=\begin{bmatrix} \epsilon_1 \\ \vdots \\ \epsilon_m \end{bmatrix} \text{ and } \epsilon_i>0 \text{ are specified constants}
\end{align}
\end{subequations}

The expression in (\ref{problem1}) can be interpreted as the weighted norm of exogenous signals, measurement noises and error in initial state estimate. It is well known that the optimal estimation algorithm for (\ref{problem1}) without $\eta_k$ is the optimal fixed interval smoother (see for example (\cite{mayne}, \cite{krener},  \cite{weinert} and \cite{nag1}). The last term above of the optimization problem (\ref{problem1}) and constraints on  $\eta_k$ imply that there there is no penalty for prediction error $(y_k-C\hat{x}_k)$ upto threshold level of $\epsilon$. Indeed, if $|y_k-C\hat{x}_k|\leq \epsilon$, the choice of $\eta_k=y_k-C\hat{x}_k$ satisfies constraint (\ref{constr2b}) while also making the last term in the  optimization problem (\ref{problem1}) zero. 

The next problem considered is the estimation with $\epsilon$- insensitive Huber loss function of the form described in (\ref{hubercost}) that also provides robustness against outliers.

\textbf{Problem 2 -} \textit{Estimation with $\epsilon$- insensitive Huber loss function :} Let P, Q be positive definite matrices and $\{r_1,\cdots, r_m\}$, $\{\kappa_1,\cdots, \kappa_m\}$ and $\{\epsilon_1,\cdots, \epsilon_m\}$ be positive scalar parameters associated with the Huber cost function. Given measurements $\{y_1, \hdots , y_N\}$, obtain the estimates of $x_k$ from the following optimization problem:
\begin{equation}
    \label{problem1h}
    \argmin_{\hat{x}_0,\hat{x}_k, \hat{w}_k}  \left[ \frac{1}{2} (\hat{x}_0-\bar{x}_0)'P(\hat{x}_0-\bar{x}_0) + \frac{1}{2} \sum_{k=0}^{N-1} \hat{w}_k'Q \hat{w}_k + \sum_{k=1}^{N} \sum_{j=1}^{m} f_{Huber}\left((y_{k}-C\hat{x}_{k})_j;r_j,\epsilon_j, \kappa_j \right)
    \right]
\end{equation}
where $(y_{k}-C\hat{x}_{k})_j$ is the $j'th$ element of $(y_{k}-C\hat{x}_{k})$, $f_{Huber}$ is the $\epsilon$- insensitive Huber loss function defined in (\ref{hubercost}) and the above optimization is subject to the constraints (\ref{constrdynamics}) describing the known system dynamics. 

The additional two problems considered in \cite{nagpalarxiv} are extensions of the above two where additional information is available about the states and exogenous signals or measurement noises that can improve estimation by identifying noisy outliers and also enhance the accuracy of the estimate by limiting possible trajectories. 
We will assume that the additional information about the system can be written as linear inequality constraints involving $x_k$ and $w_k$, as described below: 
\begin{equation}
\label{addconstr}
    \sum_{k=1}^NU_kx_k +  \sum_{k=0}^{N-1}V_kw_k \leq a \; \; \; \text{where $a \in \mathbb{R}^p$, $U_k \in \mathbb{R}^{p \times n}$ and $V_k \in \mathbb{R}^{p \times l}$ }
\end{equation}
In the above $a$, $U_k$ and $V_k$ are known matrices that may depend on the measurements $\{y_1, \cdots,y_N \}$ and each of the $p$ rows of the above inequality describes a different inequality constraint. For example if some state constraints are known in terms of average measurements such as $Lx_k \leq (b \frac{1}{N} \sum_1^N y_k + c ) $ for all $k$, we would have
\[
U_1=\begin{bmatrix} L \\ 0 \\ \vdots \\ 0 \end{bmatrix} \; , \; U_2=\begin{bmatrix} 0 \\ L \\ 0 \\ \vdots \end{bmatrix} \; , \;
\cdots \; , \; U_N=\begin{bmatrix} 0 \\  \vdots \\ 0 \\ L \end{bmatrix} , \;  \; V_k=0 \; \forall k \; , \; \text{and } \;  \;
a=\frac{1}{N} \begin{bmatrix} b \\  \vdots \\ b \end{bmatrix}  \sum_1^N y_k + \begin{bmatrix} c \\  \vdots \\ c \end{bmatrix}
\]
One also notes that though $v_k$ does not appear in the constraint described in (\ref{addconstr}), constraints involving $v_k$ can also be written as above since $v_k$ is a linear function of $x_k$ ($v_k=y_k-Cx_k$).  

The other two estimation problems of interest are extensions of the above two problems subject to the additional linear constraints (\ref{addconstr}).

\textbf{Problem 3 -} \textit{Estimation with $\epsilon$- insensitive quadratic loss function and constraints:}  Problem 1 subject to additional constraints (\ref{addconstr}).

\textbf{Problem 4 -} \textit{Estimation with $\epsilon$- insensitive Huber loss function and constraints:} Problem 2 subject to additional constraints (\ref{addconstr}).

The following definitions will be used in describing the results from \cite{nagpalarxiv}:
\begin{subequations}
\begin{align}
\label{FGdef}
  &   F  :=\begin{bmatrix} CB & 0 & \cdots & 0 \\
 CAB & CB & 0 & \cdots  \\
 \cdots & \cdots & \cdots & \cdots \\
 CA^{N-2}B & \cdots & CB & 0 \\
 CA^{N-1}B & \cdots & CAB & CB \end{bmatrix} \;  ,  \;
Y  :=\begin{bmatrix} y_1-CA\bar{x}_0 \\ \vdots \\ y_N-CA^N\bar{x}_0 \end{bmatrix} \; , \;
Q_{inv}:= \text{diag}(\{Q^{-1}\}) \in  \mathbb{R}^{(N-1)l \times (N-1)l}
  \\
 \label{XYdef}
 &   R_{inv}:= \text{diag}(\{R^{-1}\}) \ \in  \mathbb{R}^{Nm \times Nm} 
  \; , \; R_{Hinv}:= \text{diag}(\{R_H^{-1}\}) \  \in  \mathbb{R}^{Nm \times Nm} \; \text{where } \ R_H= \text{diag}(r_i) \in \mathbb{R}^{m \times m} \\
      \label{epskappa}
     & \varepsilon:=\begin{bmatrix} I_m \\ \vdots \\ I_m \end{bmatrix} \begin{bmatrix} \epsilon_1  \\ \vdots \\ \epsilon_m \end{bmatrix} \; , \; 
     K:=\begin{bmatrix} I_m \\ \vdots \\ I_m \end{bmatrix} \begin{bmatrix} \kappa_1  \\ \vdots \\ \kappa_m \end{bmatrix}  \\
     \label{Mdef}
     &   M := FQ_{inv}F' + R_{inv} + \begin{bmatrix} CA \\ \vdots \\ CA^{N} \end{bmatrix} P^{-1}\begin{bmatrix} A'C' & \hdots & A'^{N}C' \end{bmatrix} \\
      \label{MHdef}
     &   M_H := FQ_{inv}F' + R_{Hinv} + \begin{bmatrix} CA \\ \vdots \\ CA^{N} \end{bmatrix} P^{-1}\begin{bmatrix} A'C' & \hdots & A'^{N}C' \end{bmatrix}
 \end{align}
 \end{subequations}

\begin{equation}
    \label{VGHdef}
    V:=\begin{bmatrix} V_0 & \cdots & V_{N-1} \end{bmatrix}, \; 
    G  :=\begin{bmatrix} B' & B'A' & \cdots & B'A'^{N-1} \\ 0 & B' & \cdots & B'A'^{N-2} \\ \cdots & \cdots & \cdots & \cdots \\
 0 & \cdots & 0 & B' \end{bmatrix} \begin{bmatrix} U_1' \\ \vdots \\ U_N' \end{bmatrix},
 H:= \begin{bmatrix} CA \\ \vdots \\ CA^N \\   -\sum_{i=1}^{N} U_iA^i \end{bmatrix} 
\end{equation}
\begin{align}
\label{Tdef}
    T:=\begin{bmatrix} F \\ -(G'+V) \end{bmatrix} Q_{inv} \begin{bmatrix} F' & -(G+V') \end{bmatrix} + \begin{bmatrix}
    R_{inv} & 0 \\ 0 & 0 \end{bmatrix}+HP^{-1}H' \\
    \label{THdef}
    T_H:=\begin{bmatrix} F \\ -(G'+V) \end{bmatrix} Q_{inv} \begin{bmatrix} F' & -(G+V') \end{bmatrix} +\begin{bmatrix}
    R_{Hinv} & 0 \\ 0 & 0 \end{bmatrix}+ HP^{-1}H'
\end{align}
 
The following results from \cite{nagpalarxiv} describe optimal smoothing algorithms for $\epsilon$- insensitive quadratic loss function and $\epsilon$- insensitive Huber cost functions for the four problems described above.

 \begin{theorem}
\label{theorem1}
(Problem $1$) : 
Let $\hat{\Theta} , \hat{\zeta} \in \mathbb{R}^{Nm}$ be the solution of the following quadratic optimization problem with linear constraints:

\begin{subequations}
\begin{align}
\label{opteps}
& \argmax_{\Theta, \zeta} \left[ - \frac{1}{2}  \Theta' M \Theta - \varepsilon \zeta  + \Theta' Y  \right] \\
\label{constreps}
& \text{subject to constraints  } \zeta \geq \Theta \text{  and  } \zeta \geq -\Theta  
\end{align}
\end{subequations}

Let $\lambda_k$ be obtained from optimal $\hat{\Theta}$ as follows:

\begin{equation}
\label{optlambdaep}
\lambda_{k-1}  = A'\lambda_{k}+C'\hat{\theta}_k \text{ , } \lambda_N=0 \; \text{   where   }    \begin{bmatrix} \hat{\theta}_1 \\ \vdots \\ \hat{\theta}_N \end{bmatrix}=\hat{\Theta}  \text{    and   } \hat{\theta}_k \in \mathbb{R}^m
\end{equation}
With $\lambda_k$ obtained as above, the optimal values for $\hat{x}_k$, $\hat{w}_k$ that minimize (\ref{problem1}) subject to constraints (\ref{constrdynamics}) and (\ref{constr2b}) are obtained as follows:
\begin{subequations}
\begin{align}
\label{optxpr1}
& \hat{x}_{k+1}  =  A\hat{x}_k+BQ^{-1}B'\lambda_k \text{ , } \hat{x}_0= \bar{x}_0 +P^{-1}A'\lambda_0 \\
& \hat{w}_k=Q^{-1}B'\lambda_k 
\end{align}
 \end{subequations}
 \end{theorem}

\begin{theorem}
\label{theorem2}
(Problem $2$): Let $\hat{\Theta} , \hat{\zeta} \in \mathbb{R}^{Nm}$ be the solution of the following quadratic optimization problem with linear constraints:

\begin{subequations}
\begin{align}
\label{optepsH}
& \argmax_{\Theta, \zeta} \left[ - \frac{1}{2}  \Theta' M_H \Theta   - \varepsilon \zeta  + \Theta' Y  \right] \\
\label{constrepsH}
& \text{subject to constraints  } \zeta \geq \Theta, \; \zeta \geq -\Theta  , \; \Theta \leq K \text{  and  }  -\Theta  \leq K
\end{align}
\end{subequations}

where $\varepsilon$ and $K$ are as defined in (\ref{epskappa}). Let $\lambda_k$ be obtained from optimal $\hat{\Theta}$ as follows:

\begin{equation}
\label{optlambdaepH}
\lambda_{k-1}  = A'\lambda_{k}+C'\hat{\theta}_k \text{ , } \lambda_N=0 \; \text{   where   }    \begin{bmatrix} \hat{\theta}_1 \\ \vdots \\ \hat{\theta}_N \end{bmatrix}=\hat{\Theta}  \text{    and   } \hat{\theta}_k \in \mathbb{R}^m
\end{equation}
With $\lambda_k$ obtained as above, the optimal values for $\hat{x}_k$, $\hat{w}_k$ that minimize (\ref{problem1h}) subject to constraint  (\ref{constrdynamics}) are obtained as follows:
\begin{subequations}
\begin{align}
\label{optxpr1H}
& \hat{x}_{k+1}  =  A\hat{x}_k+BQ^{-1}B'\lambda_k \text{ , } \hat{x}_0= \bar{x}_0 +P^{-1}A'\lambda_0 \\
& \hat{w}_k=Q^{-1}B'\lambda_k 
\end{align}
 \end{subequations}
 \end{theorem}

 \begin{theorem}
\label{theorem3}
(Problem 3): Let $\hat{\Theta} , \hat{\zeta} \in \mathbb{R}^{Nm}$ and $\hat{\xi} \in \mathbb{R}^{p}$ be the solution of the following quadratic optimization problem with linear constraints:

\begin{subequations}
\begin{align}
\label{opteps2}
\argmax_{\Theta, \zeta, \xi} & \left[ - \frac{1}{2} \begin{bmatrix} \Theta & \xi \end{bmatrix}' T \begin{bmatrix} \Theta \\ \xi \end{bmatrix} 
-  \varepsilon' \zeta + \Theta' Y - \xi'(a-\sum_{i=1}^NU_iA^i\bar{x}_0) \right] \\
\label{constreps2}
& \text{subject to constraints  } \zeta \geq \Theta \text{  ,  } \zeta \geq -\Theta  \; \; \text{and } \xi \geq0
\end{align}
\end{subequations}

Let $\lambda_k$ be obtained from optimal $\hat{\Theta}$ and $\hat{\xi}$ as follows

\begin{equation}
\label{optlambdaep2}
\lambda_{k-1}  = A'\lambda_{k}+C'\hat{\theta}_k -U_k' \hat{\xi} \; \text{ , } \lambda_N=0 \; \text{   where   }    \begin{bmatrix} \hat{\theta}_1 \\ \vdots \\ \hat{\theta}_N \end{bmatrix}=\hat{\Theta}  \text{    and   } \hat{\theta}_k \in \mathbb{R}^m
\end{equation}
With $\lambda_k$ obtained as above, the optimal values for $\hat{x}_k$, $\hat{w}_k$ that minimize (\ref{problem1}) subject to constraints (\ref{constrdynamics}), (\ref{constr2b}) and (\ref{addconstr}) are obtained as follows:
\begin{subequations}
\begin{align}
\label{optxpr2}
& \hat{x}_{k+1}  =  A\hat{x}_k+BQ^{-1}B'\lambda_k -BQ^{-1}V_k' \hat{\xi} \; , \; \; \hat{x}_0= \bar{x}_0 +P^{-1}A'\lambda_0 \\
& \hat{w}_k=Q^{-1}B'\lambda_k - Q^{-1}V_k' \hat{\xi} 
\end{align}
 \end{subequations}
 \end{theorem}

  \begin{theorem}
\label{theorem4}
(Problem 4): Let $\hat{\Theta} , \hat{\zeta} \in \mathbb{R}^{Nm}$ and $\hat{\xi} \in \mathbb{R}^{p}$ be the solution of the following quadratic optimization problem with linear constraints:

\begin{subequations}
\begin{align}
\label{opteps2H}
\argmax_{\Theta, \zeta, \xi} & \left[ - \frac{1}{2} \begin{bmatrix} \Theta & \xi \end{bmatrix}' T_H \begin{bmatrix} \Theta \\ \xi \end{bmatrix} 
-  \varepsilon' \zeta + \Theta' Y - \xi'(a-\sum_{i=1}^NU_iA^i\bar{x}_0) \right] \\
\label{constreps2H}
& \text{subject to constraints  } \zeta \geq \Theta \text{  ,  } \zeta \geq -\Theta . \; \Theta \leq K, \;  -\Theta \leq K, \text{and } \xi \geq0
\end{align}
\end{subequations}

where $\varepsilon$ and $K$ are as defined in (\ref{epskappa}). Let $\lambda_k$ be obtained from optimal $\hat{\Theta}$ and $\hat{\xi}$ as follows:

\begin{equation}
\label{optlambdaep2H}
\lambda_{k-1}  = A'\lambda_{k}+C'\hat{\theta}_k -U_k' \hat{\xi} \; \text{ , } \lambda_N=0 \; \text{   where   }    \begin{bmatrix} \hat{\theta}_1 \\ \vdots \\ \hat{\theta}_N \end{bmatrix}=\hat{\Theta}  \text{    and   } \hat{\theta}_k \in \mathbb{R}^m
\end{equation}
With $\lambda_k$ obtained as above, the optimal values for $\hat{x}_k$, $\hat{w}_k$ that minimize (\ref{problem1h}) subject to constraints (\ref{constrdynamics}) and (\ref{addconstr}) are obtained as follow:
\begin{subequations}
\begin{align}
\label{optxpr2H}
& \hat{x}_{k+1}  =  A\hat{x}_k+BQ^{-1}B'\lambda_k -BQ^{-1}V_k' \hat{\xi} \; , \; \; \hat{x}_0= \bar{x}_0 +P^{-1}A'\lambda_0 \\
& \hat{w}_k=Q^{-1}B'\lambda_k - Q^{-1}V_k' \hat{\xi} \nonumber
\end{align}
 \end{subequations}
 \end{theorem}

 \section{Main Results}
 
The estimation algorithms described in the previous section are not recursive and require one to solve a quadratic optimization problem where the size of quadratic optimization grows linearly with $N$, the number of observations. Here we will apply the previous results to one time step to obtain recursive filtering algorithms that bear close structural similarity with Kalman filter. 

Let us assume we are at time step $k$ and have obtained estimate of state $\hat{x}_k$ based on measurements $\{y_1,\cdots,y_k\}$. Our objective is to estimate $\hat{x}_{k+1}$ for the dynamical system (\ref{system}) given the new measurement $y_{k+1}$ so that the estimate $\hat{x}_{k+1}$ minimizes the following:

\begin{equation}
    \label{problem1f}
    \argmin_{\hat{\hat{x}}_{k}, \hat{w}_k, \eta_k}  \left[ (\hat{\hat{x}}_k-\hat{x}_k)'P(\hat{\hat{x}}_k-\hat{x}_k) +  \hat{w}_k'Q \hat{w}_k + (y_{k+1}-C\hat{x}_{k+1}-\eta_k)'R(y_{k+1}-C\hat{x}_{k+1}-\eta_k) \right]
\end{equation}
subject to constraints on $\eta_k$ and system dynamics:
\begin{subequations}
\begin{align} 
\label{constrdynamicsf1}
\mid \eta_k \mid & \leq \epsilon \\
\label{constrdynamicsf2}
\hat{x}_{k+1} & =  A\hat{\hat{x}}_k+B\hat{w}_k  
\end{align}
\end{subequations}
In the above optimization problem (\ref{problem1f}), $\hat{\hat{x}}_k$ should be viewed as the updated estimate of $x_k$ after $y_{k+1}$ is observed ) (\textit{a posteriori} estimate of $x_k$) while $\hat{x}_k$ is the prior estimate based on measurements only up to time $k$, i.e. $\{y_1, \cdots, y_{k}\}$. The following result describes the algorithm to estimate $\hat{x}_{k+1}$ from $\hat{x}_{k}$ and $y_{k+1}$ that minimizes (\ref{problem1f}) subject to constraints (\ref{constrdynamicsf1}) and (\ref{constrdynamicsf2}). 

\begin{theorem}
\label{theorem1f}
(Problem $1$) : 
Let $\hat{x}_k$ be the estimate of $x_k$ at time $k$ based on measurements up to time $k$ and let $\hat{\theta} , \hat{\zeta} \in \mathbb{R}^{m}$ be the optimal values of $\theta$ and $\zeta$ for the following quadratic optimization problem with linear constraints:

\begin{subequations}
\begin{align}
\label{optepsf}
& \argmax_{\theta, \zeta} \left[ - \frac{1}{2}  \theta' (CBQ^{-1}B'C'+R^{-1}+CAP^{-1}A'C') \theta - \epsilon' \zeta  + \theta' (y_{k+1}-CA\hat{x}_k)  \right] \\
\label{constrepsf}
& \text{subject to constraints  } \zeta \geq \theta \text{  and  } \zeta \geq -\theta  
\end{align}
\end{subequations}

Then the optimal estimate of $x_{k+1}$ that minimizes (\ref{problem1f}) subject to constraints (\ref{constrdynamicsf1}) and (\ref{constrdynamicsf2}) is obtained as follows

\begin{equation}
\label{xestPr1}
\hat{x}_{k+1}= A\hat{x}_k+(AP^{-1}A'+BQ^{-1}B')C'\hat{\theta}
\end{equation}
 \end{theorem}

\textbf{Proof : } Note that the optimization problem (\ref{problem1f}) is the same as that in (\ref{problem1}) with $N=1$ and $\bar{x}_0$ and $\hat{x}_0$ replaced by $\hat{x}_k$ and $\hat{\hat{x}}_k$. We next show that the recursive algorithm (\ref{xestPr1}) follows from Theorem \ref{theorem1} with $N=1$. From (\ref{optlambdaep}), (\ref{optxpr1}) one observes
\begin{subequations}
\begin{align}
\lambda_{k}  & = A'\lambda_{k+1} + C'\hat{\theta} \; \text{ where  } \lambda_{k+1}=0, \ \text{ which implies } \ \lambda_{k}=C'\hat{\theta} \nonumber \\
\hat{x}_{k+1}  & =  A\hat{\hat{x}}_k+BQ^{-1}B'\lambda_k \; , \text{ where } \hat{\hat{x}}_k = \hat{x}_k  +P^{-1}A'\lambda_k \nonumber \\ 
\end{align}
\end{subequations}
The recursive estimate of $\hat{x}_{k+1}$ in (\ref{xestPr1}) follows from the above after substituting for $\lambda_{k}$ and $\hat{\hat{x}}_k$. $\square$

\textit{Remark 1 - Structural Similarities with Kalman Filter :} To show structural similarities with Kalman filter consider the filtering problem for system (\ref{system}) where $w_k$ and $v_k$ are independent zero mean white noises and $P^{-1}$ is the steady state variance of the state estimation error with with the following summarizing the covariances:
 \begin{equation}
 \label{KFcov}
     E\{ w_k w_j'\}=Q^{-1} \delta_{kj} \ , \; E\{v_k v_j'\}=R^{-1} \delta_{kj} \ , \; E\{(x_k-\hat{x}_k)(x_k-\hat{x}_k)'\}=P^{-1}
 \end{equation}
  
 With the above variances of $w_k$ and $v_k$, the Kalman Filter algorithm is
 \[
 \hat{x}_{k+1}= A\hat{x}_k+(AP^{-1}A'+BQ^{-1}B')C'(CBQ^{-1}B'C'+R^{-1}+CAP^{-1}A'C')^{-1}(y_{k+1}-CA\hat{x}_k)
 \]
Notice that the above is the same as (\ref{xestPr1}) if $\hat{\theta}=(CBQ^{-1}B'C'+R^{-1}+CAP^{-1}A'C')^{-1}(y_{k+1}-CA\hat{x}_k)$. If $\epsilon=0$, it can easily be seen that this is indeed the optimal $\hat{\theta}$ for the optimization problem (\ref{optepsf}). Thus $\hat{\theta}$ should be viewed as the transformed version of the innovation term $(y_{k+1}-CA\hat{x}_k)$.

\textit{Remark 2 - $\epsilon$ insensitivity to the new information ($y_{k+1}-CA\hat{x_k}$) in the estimation update :} To intuitively understand the nature of update in the above algorithm relative to Kalman filer, consider the case when $y_k$ and thus $\theta_k$ is a scalar. Then from the optimization problem (\ref{optepsf}) it can be seen that
\begin{subequations}
\begin{align}
& \text{if } \mid(y_{k+1}-CA\hat{x}_k) \mid \leq \epsilon \; \text{ then } \hat{\theta}  =0  \nonumber \\
& \text{if } (y_{k+1}-CA\hat{x}_k) \geq \epsilon \; \text{ then } \hat{\theta}  = (CBQ^{-1}B'C'+R^{-1}+CAP^{-1}A'C')^{-1}(y_{k+1}-CA\hat{x}_k -\epsilon)  \nonumber \\
& \text{if } (y_{k+1}-CA\hat{x}_k) \leq -\epsilon \; \text{ then } \hat{\theta}  = (CBQ^{-1}B'C'+R^{-1}+CAP^{-1}A'C')^{-1}(y_{k+1}-CA\hat{x}_k + \epsilon)  \nonumber 
\end{align}
\end{subequations}
In the first case if the innovation term is within a tube of $\epsilon$ (i.e., $\mid (y_{k+1}-CA\hat{x}_k) \mid \leq \epsilon$) one notes from (\ref{xestPr1}) that $\hat{x}_{k+1}= A\hat{x}_k$. Thus the algorithm chooses the prior estimate and does not incorporate $y_{k+1}$ in estimating $x_{k+1}$ if the innovation term $(y_{k+1}-CA\hat{x}_k)$ is sufficiently small. When innovation term is large and positive, i.e. $(y_{k+1}-CA\hat{x}_k) \geq \epsilon$, then from the expression of $\hat{\theta}_k$ one notes that 
\[
 \hat{x}_{k+1}= A\hat{x}_k+(AP^{-1}A'+BQ^{-1}B')C'(CBQ^{-1}B'C'+R^{-1}+CAP^{-1}A'C')^{-1}(y_{k+1}-CA\hat{x}_k-\epsilon)
 \]
Notice the similarity with Kalman filter in the previous remark with the difference being the innovation term is reduced by $\epsilon$ in the estimation update, or equivalently the results would match that of Kalman Filter if the measurement $y_{k+1}$ was replaced by $(y_{k+1}-\epsilon)$ and the weighting matrices $P$, $Q$ and $R$ were variances defined in (\ref{KFcov}). Similar comment applies for the third case when $(y_{k+1}-CA\hat{x}_k) \leq -\epsilon$ in which case the estimate from the above algorithm would match the Kalman Filter estimate if $y_{k+1}$ was replaced by $(y_{k+1}+\epsilon)$. 

We next consider the estimation with $\epsilon$- insensitive Huber loss function of the form described in (\ref{hubercost}). Our objective is to estimate $\hat{x}_{k+1}$ for the dynamical system (\ref{system}) given the new measurement $y_{k+1}$ so that the estimate $\hat{x}_{k+1}$ minimizes the following:

\begin{equation}
    \label{problem1hf}
    \argmin_{\hat{\hat{x}}_{k}, \hat{w}_k}  \left[ \frac{1}{2} (\hat{\hat{x}}_k-\hat{x}_k)'P(\hat{\hat{x}}_k-\hat{x}_k) + \frac{1}{2}  \hat{w}_k'Q \hat{w}_k + \sum_{j=1}^{m} f_{Huber}\left((y_{k}-C\hat{x}_{k})_j;r_j,\epsilon_j, \kappa_j  \right) \right]
\end{equation}
subject to the constraints on system dynamics:
\[
\hat{x}_{k+1}  =  A\hat{\hat{x}}_k+B\hat{w}_k  
\]

\begin{theorem}
\label{theorem2f}
(Problem $2$) : 
Let $\hat{x}_k$ be the estimate of $x_k$ at time $k$ based on measurements up to time $k$ and let $\hat{\theta} , \hat{\zeta} \in \mathbb{R}^{m}$ be the solution of the following quadratic optimization problem with linear constraints:

\begin{subequations}
\begin{align}
\label{optepsHf}
& \argmax_{\theta, \zeta} \left[ - \frac{1}{2}  \theta' (CBQ^{-1}B'C'+R^{-1}+CAP^{-1}A'C') \theta - \epsilon' \zeta  + \theta' (y_{k+1}-CA\hat{x}_k)  \right] \\
\label{constrepsHf}
& \text{subject to constraints  } \zeta \geq \theta \text{  and  } \zeta \geq -\theta  \; , \; \theta \leq \kappa \text{  and  }  -\theta  \leq \kappa
\end{align}
\end{subequations}

Then the optimal estimate of $x_{k+1}$ that minimizes (\ref{problem1hf}) subject to constraints (\ref{constrdynamicsf2}) is obtained as follows

\begin{equation}
\label{xestPr2}
\hat{x}_{k+1}= A\hat{x}_k+(AP^{-1}A'+BQ^{-1}B')C'\hat{\theta}
\end{equation}
 \end{theorem}

\textbf{Proof : } Similar to the previous case, the result follows from Theorem \ref{theorem2f} with $N=1$ and $\bar{x}_0$ and $\hat{x}_0$ replaced by $\hat{x}_k$ and $\hat{\hat{x}}_k$. $\square$

\textit{Remark 3 - Lower sensitivity to outliers in the Theorem \ref{theorem2f} algorithm:} The estimation algorithms for both Theorems \ref{theorem1f} and \ref{theorem2f} are similar except for the additional constraint of $|\theta| \leq \kappa$. As discussed in Remark 2 above, $\hat{\theta}$ is a scaled version of the innovation term $(y_{k+1}-CA\hat{x}_k)$. If $y_{k+1}$ is an outlier, or equivalently $\mid(y_{k+1}-CA\hat{x}_k)\mid$ is very large, this constraint on $|\theta|$ acts to lower the sensitivity of outlier $y_{k+1}$ on the estimate $\hat{x}_{k+1}$.

For the filtering problem with state and exogenous constraints, similar to (\ref{addconstr}) we will incorporate the following constraint at each time step:
\begin{equation}
\label{addconstrf}
    Ux_{k+1} +  Vw_k \leq a \; \; \text{where 
 $U, V$ and $a$ can be time dependent and are assumed to be known at time $k+1$}
\end{equation}
While these constraints are written in terms of $x_{k+1}$ and $w_k$, one notes that they can also be used to describe constraints on $v_{k+1}$. For example, the constraint on measurement noise $|v_k| \leq b$ can be written as $-Cx_k  \leq b-y_k \text{ , } Cx_k \leq b +y_k$. 

For the following result we will use the following notation:
\begin{equation}
\label{Tf}
    T_f:= \begin{bmatrix} CB \\ -UB-V \end{bmatrix} Q^{-1} \begin{bmatrix} B'C' & -B'U^{'}-V^{'} \end{bmatrix}
\end{equation}

\begin{theorem}
\label{theorem3f}
(Problem $3$) : 
Let $\hat{x}_k$ be the estimate of $x_k$ at time $k$ based on measurements up to time $k$ and let $\hat{\theta}$ and $\hat{\xi}$ be the optimal $\theta$ and $\xi$ for the following quadratic optimization problem with linear constraints:

\begin{subequations}
\begin{align}
\label{optepsHf}
& \argmax_{\theta, \xi, \zeta} \left[ - \frac{1}{2} \begin{bmatrix} \theta & \xi \end{bmatrix}' T_f \begin{bmatrix} \theta \\ \xi \end{bmatrix} 
-  \epsilon' \zeta + \theta' (y_{k+1}-CA\hat{x}_k) - \xi'(a-UA\hat{x}_k) \right] \\
\label{constrepsHf}
& \text{subject to constraints  } \zeta \geq \theta \text{  and  } \zeta \geq -\theta \; , \; \text{  and  }  \xi  \geq 0
\end{align}
\end{subequations}

Then the optimal estimate of $x_{k+1}$ that minimizes (\ref{problem1f}) subject to constraints (\ref{constrdynamicsf2}) and (\ref{addconstrf}) is obtained as follows

\begin{equation}
\label{xestPr3}
\hat{x}_{k+1}= A\hat{x}_k+(AP^{-1}A'+BQ^{-1}B')(C'\hat{\theta}-U^{'}\hat{\xi})-BQ^{-1}V^{'}\hat{\xi}
\end{equation}
 \end{theorem}

\textbf{Proof : } As in the first result above, note that the optimization problem (\ref{problem1f}) is the same as that in (\ref{problem1}) with $N=1$ and $\bar{x}_0$ and $\hat{x}_0$ replaced by $\hat{x}_k$ and $\hat{\hat{x}}_k$. We next show that the recursive algorithm (\ref{xestPr3}) follows from Theorem \ref{theorem3} with $N=1$. From (\ref{optlambdaep2}), (\ref{optxpr2}) one observes
\begin{subequations}
\begin{align}
\lambda_{k}  & = \lambda_{k-1}  = A'\lambda_{k}+C'\hat{\theta}_k -U' \hat{\xi}  \; \text{ where  } \lambda_{k+1}=0, \ \text{ which implies } \ \lambda_{k}=C'\hat{\theta} -U' \hat{\xi} \nonumber \\
\hat{x}_{k+1}  & =  A\hat{\hat{x}}_k+BQ^{-1}B'\lambda_k - BQ^{-1}V'\hat{\xi} \; , \text{ where } \hat{\hat{x}}_k = \hat{x}_k  +P^{-1}A'\lambda_k \nonumber \\ 
\end{align}
\end{subequations}
The recursive estimate of $\hat{x}_{k+1}$ in (\ref{xestPr3}) follows from the above after substituting for $\lambda_{k}$ and $\hat{\hat{x}}_k$. $\square$

The next result describes estimation algorithm for minimizing Huber cost function (\ref{problem1hf}) subject to constraints (\ref{constrdynamicsf2}) and (\ref{addconstrf}).

\begin{theorem}
\label{theorem4f}
(Problem $4$) : 
Let $\hat{x}_k$ be the estimate of $x_k$ at time $k$ based on measurements up to time $k$ and let $\hat{\theta}$ and $\hat{\xi}$ be the optimal $\theta$ and $\xi$ for the following quadratic optimization problem with linear constraints:

\begin{subequations}
\begin{align}
\label{optepsHf}
& \argmax_{\theta, \xi, \zeta} \left[ - \frac{1}{2} \begin{bmatrix} \theta & \xi \end{bmatrix}' T_f \begin{bmatrix} \theta_k \\ \xi \end{bmatrix} 
-  \epsilon' \zeta + \theta' (y_{k+1}-CA\hat{x}_k) - \xi'(a_k-UA\hat{\hat{x}}_k) \right] \\
\label{constrepsHf}
& \text{subject to constraints  } \zeta \geq \theta \;, \;  \zeta \geq -\theta \; , \; \theta \leq \kappa \; , \;  -\theta  \leq \kappa\; \text{  and  }  \xi  \geq 0
\end{align}
\end{subequations}

Then the optimal estimate of $x_{k+1}$ that minimizes (\ref{problem1hf}) subject to constraints (\ref{constrdynamicsf2}) and (\ref{addconstrf}) is obtained as follows

\begin{equation}
\label{xestPr4}
\hat{x}_{k+1}= A\hat{x}_k+(AP^{-1}A'+BQ^{-1}B')(C'\hat{\theta}-U^{'}\hat{\xi})-BQ^{-1}V^{'}\hat{\xi}
\end{equation}
 \end{theorem}

 \textbf{Proof : } Similar to the previous case, this result follows from Theorem \ref{theorem4} with $N=1$ and $\bar{x}_0$ and $\hat{x}_0$ replaced by $\hat{x}_k$ and $\hat{\hat{x}}_k$. $\square$

\textit{Remark 4 - Lower sensitivity to outliers in the algorithm of Theorem \ref{theorem4f} relative to \ref{theorem3f}:} Similar to the Remark 3, the difference between the estimation algorithms in Theorems \ref{theorem3f} and \ref{theorem4f} is the additional constraint of $|\theta| \leq \kappa$. As discussed in Remark 2 above, $\hat{\theta}$ is a scaled version of the innovation term $(y_{k+1}-CA\hat{x}_k)$ and a cap on $\hat{\theta}$ puts a constraint on maximum impact of an outlier observation of $y_{k+1}$ on the estimate $\hat{x}_{k+1}$.

\section{Summary}
This paper presents recursive estimation algorithms for linear dynamical systems with two cost functions that may be suitable for situations when measurements are biased and/or have occasional outliers. The first approach with $\epsilon$-insensitive loss functions is often used in Support Vector Machines and provides greater robustness and lower sensitivity to measurement noises as small errors are ignored. The second cost functions for which filtering algorithms are developed is the $\epsilon$-insensitive Huber cost function for which the penalty function switches from quadratic to linear for large errors which makes the estimates less sensitive to outliers. We also present algorithms to estimate states with the same objective functions while also incorporating additional constraints about the states and exogenous signals. In all cases, the algorithms are similar to Kalman-Bucy Filter and easily implemented in a recursive manner. Compared to Kalman-Bucy filter, the proposed algorithms require an additional step of obtaining solution of a quadratic optimization problems with linear constraints.

\bibliographystyle{unsrt}  

\begin{thebibliography}{1}

\bibitem{aless}
Alessandri A and Awawdeh M. 
\newblock Moving-horizon estimation with guaranteed robustness for discrete-time linear systems and measurements subject to outliers.
\newblock  Automatica, 2016.

\bibitem{anderson}
Anderson BDO and Moore JB. 
\newblock Optimal Filtering.
\newblock  Upper Saddle River, Prentice-Hall, 1979.

\bibitem{andri}
Andrien ARP and Antunes D. 
\newblock Filtering and smoothing in the presence of outliers using duality and relaxed dynamic programming.
\newblock  IEEE Conference on Decision and Control, 2019.

\bibitem{aravkin}
Aravkin A, Burke JV, Ljung L, Lozano A, and Pillonetto G. 
\newblock Generalized Kalman Smoothing: Modeling and Algorithms.
\newblock  Automatica, Volume 86, 2017.

\bibitem{chan}
Chan SC, Zhang Z and Tse KW.
\newblock A new robust Kalman filter algorithm under outliers and
system uncertainties.
\newblock {\em IEEE International Symposium on Circuits and Systems
Proceedings}, Volume 5, 2005.


\bibitem{durovic}
Durovic ZM and Kovacevic BD. 
\newblock Robust Estimation with Unknown Noise Statistics.
\newblock {\em IEEE Transactions on Automatic Control}, Volume 44, Issue 6, 1999.

\bibitem{grimble}
Elsayed A and Grimble MJ. 
\newblock A new approach to $\mathcal{H}_\infty$ design of optimal digital linear filters.
\newblock {\em IMA J. Math. Contr. Informat.}, Volume 6, Issue 8, 1989.

\bibitem{garulli}
Garulli A, Vicino AA and Zappa G. 
\newblock Conditional central algorithms for worst case set-membership identification and filtering.
\newblock {\em IEEE Transactions on Automatic Control}, Volume 45, Issue 1, 2000.

\bibitem{Huber}
Huber PJ, \newblock Robust Statistics. New York: John Wiley, 1981.

\bibitem{krener}
Krener AJ. 
\newblock Kalman-Bucy and minimax filtering.
\newblock {\em IEEE Transactions on Automatic Control}, Volume 25, Issue 2, 1980.


\bibitem{meditch}
Meditch JS.
\newblock Stochastic optimal linear estimation and control.
\newblock {\em McGraw-Hill}, 1969.


\bibitem{milanese}
Milanese M and Vicino A. 
\newblock Optimal estimation theory for dynamic systems with set membership uncertainty : an overview.
\newblock {\em Automatica}, Volume 26 (1991), Issue 6.


\bibitem{nagpalarxiv}
Nagpal KM.
\newblock Estimation of Dynamical Systems in Noisy Conditions and with Constraints. 	arXiv:2011.02648 , (2022).

\bibitem{nag1}
Nagpal KM, and Khargonekar PP.
\newblock Filtering and Smoothing in an $\mathcal{H}_\infty$ setting.
\newblock {\em IEEE Transactions on Automatic Control}, Volume 36 (1991), Issue 2.

\bibitem{pontil}
Pontil M, Mukherjee S and Girosi F. 
\newblock On the noise model of support vector machines regression.
\newblock {\em Algorithmic Learning Theory}, Springer, 2000.


\bibitem{rojo}
Rojo-Alvarez JL, Martinez-Ramon M, de Prado-Cumplido M, Artes-Rodriguez A and Figueiras-Vidal AR.
\newblock Support Vector Method for Robust ARMA System Identification.
\newblock {\em IEEE Transactions on Signal Processing}, Volume 52 (2004), No. 1.

\bibitem{shaked}
Shaked U.
\newblock $\mathcal{H}_\infty$ Control Minimum Error State Estimation of linear stationary processes.
\newblock {\em IEEE Transactions on Automatic Control}, Vol 35 (1990), Issue 5.



\bibitem{vapnik1}
Vapnik V, Golowich S and Smola A.
\newblock Support vector method for function approximation, regression, estimation, and signal processing.
\newblock {\em Advances in Neural Information Processing Systems 9}, pages 81-287, MIT Press, 1997.

\bibitem{vapnik2}
Vapnik V.
\newblock Statistical Learning Theory.
\newblock {\em John Wiley \& Sons}, New York, 1998.





\end{thebibliography}

\end{document}